\documentclass[onecolumn,preprintnumbers,amsmath,aps]{revtex4}
\usepackage{graphicx}
\usepackage{dcolumn}
\usepackage{bm}
\usepackage{multirow}
\usepackage{amsmath}
\usepackage{amssymb}
\usepackage{tabularx}
\begin{document}
\newcommand{\kvec}{\mbox{{\scriptsize {\bf k}}}}
\newcommand{\qvec}{\mbox{{\scriptsize {\bf q}}}}
\def\eq#1{Eq.\hspace{1mm}(\ref{#1})}
\def\fig#1{Fig.\hspace{1mm}\ref{#1}}
\def\tab#1{Tab.\hspace{1mm}\ref{#1}}
\title
{The unbalanced phonon--induced superconducting state on a square lattice\\ beyond the static boundary}
\author{K. 
% GP\textcolor{red}
{A.} Szewczyk$^{\left(1\right)}$}
\email{kamila.szewczyk@ajd.czest.pl}
\author{M. W. Jarosik$^{\left(2\right)}$}
\author{A. P. Durajski$^{\left(2\right)}$}
\author{R. Szcz{\c{e}}{\'s}niak$^{\left(1,2\right)}$}
\affiliation{$^1$ Division of Theoretical Physics, Institute of Physics, 
                  Jan D{\l}ugosz University in Cz{\c{e}}stochowa, Ave. Armii Krajowej 13/15, 42-200 Cz{\c{e}}stochowa, Poland}
\affiliation{$^2$ Institute of Physics, 
                  Cz{\c{e}}stochowa University of Technology, Ave. Armii Krajowej 19, 42-200 Cz{\c{e}}stochowa, Poland}
\date{\today}
\begin{abstract}

The paper presents our verification of induction of the superconducting state on a square lattice by the linear electron--phonon interaction for values of the unbalance parameter ($\gamma=\lambda_{D}/\lambda_{ND}$) less than $\gamma_{C}=0.42$. Symbols $\lambda_{D}$ and $\lambda_{ND}$ denote the values of the coupling constant in the diagonal and the non--diagonal channel of the self--energy.
Calculations were carried out using the Eliashberg equations, in which the order parameter ($\Delta_{\kvec}\left(i\omega_{n}\right)$) and the wave function renormalising factor ($Z_{\kvec}\left(i\omega_{n}\right)$) depend explicitly on the Matsubara frequency ($\omega_{n}$) and the wave vector (${\bf k}$). 
The value of $\gamma_{C}$ in the static boundary ($\Delta_{\kvec}\left(i\omega_{n}\right)\rightarrow \Delta_{\kvec}\left(i\omega_{n=1}\right)$), equal to ($0.93$), is significantly greater than the obtained limit value of $0.42$. Values of the thermodynamic functions of the superconducting state determined for our assumptions are significantly different from the values calculated in accordance with the BCS theory.  The results were obtained for the electron--phonon interaction function explicitly dependent on the momentum transfer between electron states.  
\end{abstract}
\maketitle

Physical properties of the phonon--induced superconducting state on a square lattice were determined in the present work within the Eliashberg formalism. The initial results, obtained in the limit of momentum approximation (which neglects the self--consistency with respect to Matsubara frequency), were already presented by us elsewhere \cite{Szewczyk2018A}. There we showed that the balanced superconducting state on a square lattice cannot arise in the case of the constant value of the electron--phonon coupling function ($g_{\qvec}\sim g$). By the balanced superconducting state we mean such a state, for which the coupling constant $\lambda_{D}$ in the diagonal channel of the self--energy matrix ($M_{\kvec}\left(i\omega_{n}\right)$) is equal to the coupling constant in its non--diagonal channel $\lambda_{ND}$. 
Please notice that the diagonal elements of the self--energy determine, \textit{inter alia}, the influence of the electron--phonon interaction on the value of the effective electron mass. The off--diagonal elements of the  $M_{\kvec}\left(i\omega_{n}\right)$ are directly related to the order parameter of the phase transition between the superconducting and the normal state.

At present, we know many physical mechanisms which can lead to the unbalanced phonon--induced superconducting state. Particular attention should be paid to the spin--fluctuation (paramagnon) scattering, which contributes to the effective coupling constant in the diagonal channel in the opposite way than it does in the antidiagonal channel \mbox{($s$--wave} symmetry) \cite{Rietschel1979A, Rietschel1980A, Fay1980A, Daams1981A, Fay1982A, Leavens1983A}. The unbalance of the superconducting state can also be generated by the non--conventional terms of the electron--phonon interaction, e.g. the electron--electron--phonon \cite{Szczesniak2012A} or, similarly, the hole--hole--phonon interaction \cite{Durajski2018A}. Additionally, they induce the asymmetry of the electron density of states, the pseudo--gap in the electron density of states, or the anomalous dependence of the energy gap on doping \cite{Szczesniak2017A}. These effects can be of particular importance for the correct description of the superconducting state in cuprates \cite{Bednorz1986A, Bednorz1988A, Dagotto1994A}, especially since they are observed experimentally \cite{Renner1998A, Renner1998B, Fischer2007A}. 
At the level of the conventional Eliashberg formalism, the unbalanced superconducting state arises also when the non--phonon contribution from the $d$--wave type symmetry is taken into account in the $\alpha^{2}_{\kvec}F\left(\omega\right)$ Eliashberg function modeling the electron--phonon interaction \mbox{\cite{Rieck1990A, Jiang1993A, Zasadzinski2003A}}. Nevertheless the considered model is too simple for the satisfactory explanation of even some of the anomalous properties of the superconducting state in cuprates.

The purpose of our investigation, the results of which are gathered in the present work, was to find information, as precise as possible, on the physical properties of the phonon--induced superconducting state on a square lattice. Therefore we solved the Eliasberg equations \cite{Eliashberg1960A} with respect both to the electron wave vector and to the Matsubara frequency in the fully self--consistent way. We took into consideration the electron--phonon coupling function explicitly dependent on the phonon transfer of momentum between electron states \cite{Allan2017A}. It is worth noticing that the Eliashberg equations on a square lattice are for the first time analysed in such a precise manner. 

The obtained results allowed us also to comment on the problem of the significance of the electron--phonon interaction in the context of determination of the pairing mechanism in cuprates \cite{Dagotto1994A, Szczesniak2012A}. This issue is discussed at greater length in the concluding part of the work. 
 
In its most advanced form, the analysis of the phonon--induced superconducting state is performed within the Eliashberg formalism, which is based on the self--consistent calculation of the matrix elements of the thermodynamic Green's function:
$G_{\kvec}\left(i\omega_{n}\right)=\left<\left<\Psi_{\kvec}|\Psi^{\dagger}_{\kvec}\right>\right>_{i\omega_{n}}$. 
The symbol $\Psi^{\dagger}_{\kvec}$ denotes the Nambu spinor with Hermitian coupling: $\Psi^{\dagger}_{\kvec}=\left(c^{\dagger}_{\kvec\uparrow},c_{-\kvec\downarrow}\right)$, and $\omega_{n}=\frac{\pi}{\beta}\left(2n-1\right)$ stands for the fermionic Matsubara frequency ($\beta=1/k_{B}T$). The quantity $c^{\dagger}_{\kvec\sigma}$ is the creation operator of the electron state with momentum ${\bf k}$ and spin  
$\sigma\in\{\uparrow,\downarrow\}$. The explicit form of the self-energy $M_{\kvec}\left(i\omega_{n}\right)$ is determined as a rule by application of the Hamiltonian which models the linear coupling between the electron and the phonon subsystems \cite{Frohlich1952A}: 
$H=\sum_{\kvec}\varepsilon_{\kvec}\Psi^{\dagger}_{\kvec}\tau_{3}\Psi_{\kvec}+\sum_{\qvec}\omega_{\qvec}b^{\dagger}_{\qvec}b_{\qvec}+\frac{1}{\sqrt{N}}\sum_{\kvec\qvec}g_{\qvec}\Psi^{\dagger}_{\kvec+\qvec}\tau_{3}\Psi_{\kvec}\phi_{\qvec}$, where $\varepsilon_{\kvec}$ is the electronic dispersion relation 
and $\phi_{\qvec}=b_{\qvec}+b^{\dagger}_{-\qvec}$. The quantity $b^{\dagger}_{\qvec}$ denotes the creation operator of the phonon state with momentum ${\bf q}$. The symbol $\omega_{\qvec}$ represents phonon dispersion relation ($\omega_{\qvec}=\omega_{-\qvec}$). 
The relatively easy quantum operator calculations bring about the exact result:  
$M_{\kvec}\left(i\omega_{n}\right)=\frac{1}{N}\sum_{\qvec\qvec'}g_{\qvec}g_{\qvec'}\tau_{3}
\left<\left<\Psi_{\kvec-\qvec}\phi_{\qvec}|\Psi^{\dagger}_{\kvec+\qvec'}\phi_{\qvec'}\right>\right>_{i\omega_{n}}\tau_{3}$. 
Further analysis of the problem is based on the Migdal approximation and on the Wick's theorem \cite{Migdal1958A, Fetter1971A}, by virtue of which the thermodynamic Green's function  $\left<\left<\Psi_{\kvec-\qvec}\phi_{\qvec}|\Psi^{\dagger}_{\kvec+\qvec'}\phi_{\qvec'}\right>\right>_{i\omega_{n}}$ is approximated by the product of the $G_{\kvec}\left(i\omega_{n}\right)$ function and the phonon propagator for the non--interacting phonons. 
The procedure of reaching self--consistency allows to obtain the non--linear Eliashberg equations, which determine the thermodynamics of the superconducting phase on the quantitative level. For the half-filled electon band, at the occurrence of the slight particle--hole symmetry violation, the Eliashberg equations take the form:
\begin{equation}
\label{r1}
\varphi_{\bf k}\left(i\omega_{n}\right)=\frac{1}{\beta N}\sum_{m{\bf q}}
K_{\bf q}\left(\omega_{n}-\omega_{m}\right) 
\frac{\varphi_{{\bf k}-{\bf q}}\left(i\omega_{m}\right)}{D_{{\bf k}-{\bf q}}\left(i\omega_{m}\right)},
\end{equation}
\begin{equation}
\label{r2}
Z_{\bf k}\left(i\omega_{n}\right)=1+\frac{\gamma}{\beta N}\sum_{m{\bf q}}
\frac{\omega_{m}}{\omega_{n}}
K_{\bf q}\left(\omega_{n}-\omega_{m}\right)
\frac{Z_{{\bf k}-{\bf q}}\left(i\omega_{m}\right)}{D_{{\bf k}-{\bf q}}\left(i\omega_{m}\right)},
\end{equation}
The symbol $\varphi_{\bf k}\left(i\omega _{n}\right)$ denotes the order parameter function, the order parameter being defined by the $\Delta_{\bf k}\left(i\omega_{n}\right)=\varphi_{\bf k}\left(i\omega_{n}\right)/Z_{\bf k}\left(i\omega_{n}\right)$ ratio, where $Z_{\bf k}\left(i\omega_{n}\right)$ represents the wave function renormalisation factor. The pairing kernel of the electron--phonon interaction is given by the formula: $K_{\bf q}\left(\omega_{n}-\omega_{m}\right)=2g^{2}_{\bf q}\frac{\omega_{\qvec}}{\left(\omega_{n}-\omega_{m}\right)^{2}+
\omega^{2}_{\qvec}}$. Additionally, $D_{\bf k}\left(i\omega_{n}\right)=\left(\omega_{n}Z_{\bf k}\left(i\omega _{n}\right)\right)^{2}+
\varepsilon^{2}_{{\bf k}}+\varphi^{2}_{\bf k}\left(i\omega_{n}\right)$. 
Usually the Eliashberg equations are solved in the isotropic approximation: $\varphi_{\bf k}\left(i\omega_{n}\right)\sim\varphi\left(i\omega_{n}\right)$ and
$Z_{\bf k}\left(i\omega_{n}\right)\sim Z \left(i\omega_{n}\right)$, while the pairing kernel is transformed in the following way: $K_{\bf q}\left(\omega_{n}-\omega_{m}\right)\sim 2\int^{\omega_{D}}_{0}d\omega'\frac{\alpha^{2}F\left(\omega'\right)}
{\left(\omega_{n}-\omega_{m}\right)^2+\omega'^{2}}$, where $\omega_{D}$ is the Debye phonon frequency. 
The spectral function $\alpha^{2}F\left(\omega\right)$ for a specific physical system can be determined by means of the DFT method. This approach works perfectly e.g. for the recently discovered electron--phonon hydrogen--containing superconductors \cite{Li2014A, Duan2014A, Durajski2015A, Durajski2016B, Durajski2017A, Szczesniak2018A, Kostrzewa2018A, Kruglov2018A}, characterised by the extremely high values of the critical temperature: ${\rm H_{2}S}$ ($T_{C}=150$~K for $p=150$~GPa), 
${\rm H_{3}S}$ ($T_{C}=203$~K for $p=150$~GPa \cite{Drozdov2014A, Drozdov2015A}, and ${\rm LaH_{10}}$ ($T_{C}=215$~K for $p=150$~GPa) \cite{Drozdov2019A}, ($T_{C}=260$~K for $p\in\left(180-200\right)$~GPa) \cite{Somayazulu2019A}. 

Taking into account the results reported in \cite{Szewczyk2018A}, we supplemented the \eq{r2} with the $\gamma$ parameter, which determines the degree of unbalance of the electron--phonon coupling constants in the diagonal and the non--diagonal channel of self--energy. 

We performed numerical calculations on the $N\times N$ momentum lattice, where $N=200$, and  we took into account $200$ Matsubara frequencies. Therefore 
eight million equations had to be solved. To minimize the hardware requirements, the Eliashberg equations were written in the form which explicitly expressed all symmetries of solutions with respect to the wave vector and to the Matsubara frequency. Assuming that the auxiliary symbol $f_{\kvec}\left(i\omega_{n}\right)=f_{n}\left(k_{x},k_{y}\right)$ should be understood as $\varphi$ or $Z$, one can write that: 
$f_{n}\left(k_{x},k_{y}\right)=f_{n}\left(-k_{x},k_{y}\right)=f_{n}\left(k_{x},-k_{y}\right)=f_{n}\left(-k_{x},-k_{y}\right)$, and 
$f_{n}\left(k_{x},k_{y}\right)=f_{-n+1}\left(k_{x},k_{y}\right)$. The Eliashberg equations were solved by means of the AMCIA program, which generalises the numerical procedures used in another our work \cite{Szczesniak2006A}. The AMCIA numerical environment requires introduction of the initial values of the $\varphi_{\bf k}(i\omega_{n})$ and $Z_{\bf k}(i\omega_{n})$ functions. As far as the wave function renormalising factor is concerned, we took into account the first $25$ terms of the series:
\begin{eqnarray}
Z_{m_{0}}\left( {\bf q}_{0}\right)
=1+\sum^{+\infty}_{L=1}\frac{1}{\left(\beta N\right)^{L}} \sum^{\pi}_{{\bf q}_{1}\sim{\bf q}_{L}={\bf 0}^{+}} \sum^{M}_{m_{1}\sim m_{L}=1}
\frac{\omega_{m_{L}}}{\omega_{m_{0}}}\prod^{L}_{j=1}J_{m_{j-1}m_{j}}\left({\bf q}_{j-1},{\bf q}_{j}\right)d_{m_{j}}^{-1}\left({\bf q}_{j}\right),
\end{eqnarray}
where: ${\bf q}_{j}=\left(q^{j}_{x},q^{j}_{y}\right)$ and
$J_{m_{j-1}m_{j}}\left({\bf q}_{j-1},{\bf q}_{j}\right)=
J_{m_{j-1}m_{j}}\left(q^{j-1}_{x},q^{j-1}_{y},q^{j}_{x},q^{j}_{y}\right)=
K'_{m_{j-1}m_{j}}\left(q^{j-1}_{x}+q^{j}_{x},q^{j-1}_{y}+q^{j}_{y}\right)+
K'_{m_{j-1}m_{j}}\left(q^{j-1}_{x}+q^{j}_{x},q^{j-1}_{y}-q^{j}_{y}\right)+
K'_{m_{j-1}m_{j}}\left(q^{j-1}_{x}-q^{j}_{x},q^{j-1}_{y}+q^{j}_{y}\right)+
K'_{m_{j-1}m_{j}}\left(q^{j-1}_{x}-q^{j}_{x},q^{j-1}_{y}-q^{j}_{y}\right)$, while
$K'_{m_{j-1}m_{j}}\left(q_{x},q_{y}\right)=K_{\bf q}\left(\omega_{m_{j-1}}-\omega_{m_{j}}\right)-K_{\bf q}\left(\omega_{m_{j-1}}+\omega_{m_{j}}\right)$.
Additionally: $d_{m_{j}}\left({\bf q}_{j}\right)=\omega^{2}_{m_{j}}+\varepsilon^{2}_{{\bf q}_{j}}$. 
We choose positive non-zero numbers as initial values for the order parameter function. It should be mentioned that the number of the momentum lattice points and the number of Matsubara frequencies taken into account in our calculations provide for the convergence of the obtained results (see Appendix \ref{D01}).

The energy of electron states was modeled by using the tight--binding description on the ion equilibrium positions assuming the nearest--neighbor ($t$) and next--nearest neighbor ($t'$) hopping \cite{Dagotto1994A}: \mbox{$\varepsilon_{\kvec}=-2t\left[\cos\left(k_{x}\right)+\cos\left(k_{y}\right)\right]+4t'\cos\left(k_{x}\right)\cos\left(k_{y}\right)$}, where the lattice constant is taken as unity. In the present work we assumed that energy is expressed in units of hopping integral $t$, and $t'=0.1$~$t$. 
% GP \textcolor{ForestGreen}
From the physical point of view, the low value of the $t'$ integral indicates the relatively small violation of the particle--hole symmetry. Therefore it can be assumed for the considered case that the values of the chemical potential and the energy shift functions are equal to zero. For higher values of $t'$ the full system of Eliashberg equations should be analysed according to the scheme presented in References \cite{Carbotte1990A, Radtke1994A}.
% 
% GP\textcolor{blue}
We took into account the acoustic phonons stemming from the nearest--neighbor spring: 
$\omega_{\qvec}=\omega_{0}\sqrt{2-\cos{q_{x}}-\cos{q_{y}}}$. In model calculations it was assumed that $\omega_{0}=0.15$~$t$, so that the maximum phonon frequency $\omega_{D}=2\omega_{0}=0.3$~$t$.
In the case of a square lattice, the electron--phonon matrix elements are given by \cite{Allan2017A}: 
$g_{\qvec}=g_{0}|{\bf q}|\sqrt{1/\omega_{\qvec}}$, and $g_{0}=0.031$~$t^{3/2}$. 
%GP
The value of the $g_{\qvec}$ function does not exceed $0.3$~$t$ in the considered case. Let us notice that the function $g_{\qvec}$ reproduces the results reported in \cite{Madelung1978A, Coleman2015A, Giustino2016A}, and has the structure predicted by Bloch \cite{Bloch1928A}. 
It should be noticed, however, that in a general case the electron--phonon coupling function depends explicitly also on the ${\bf k}$ vector: $g_{\kvec,\kvec+\qvec}$.

\begin{figure}
\includegraphics[width=0.3\columnwidth]{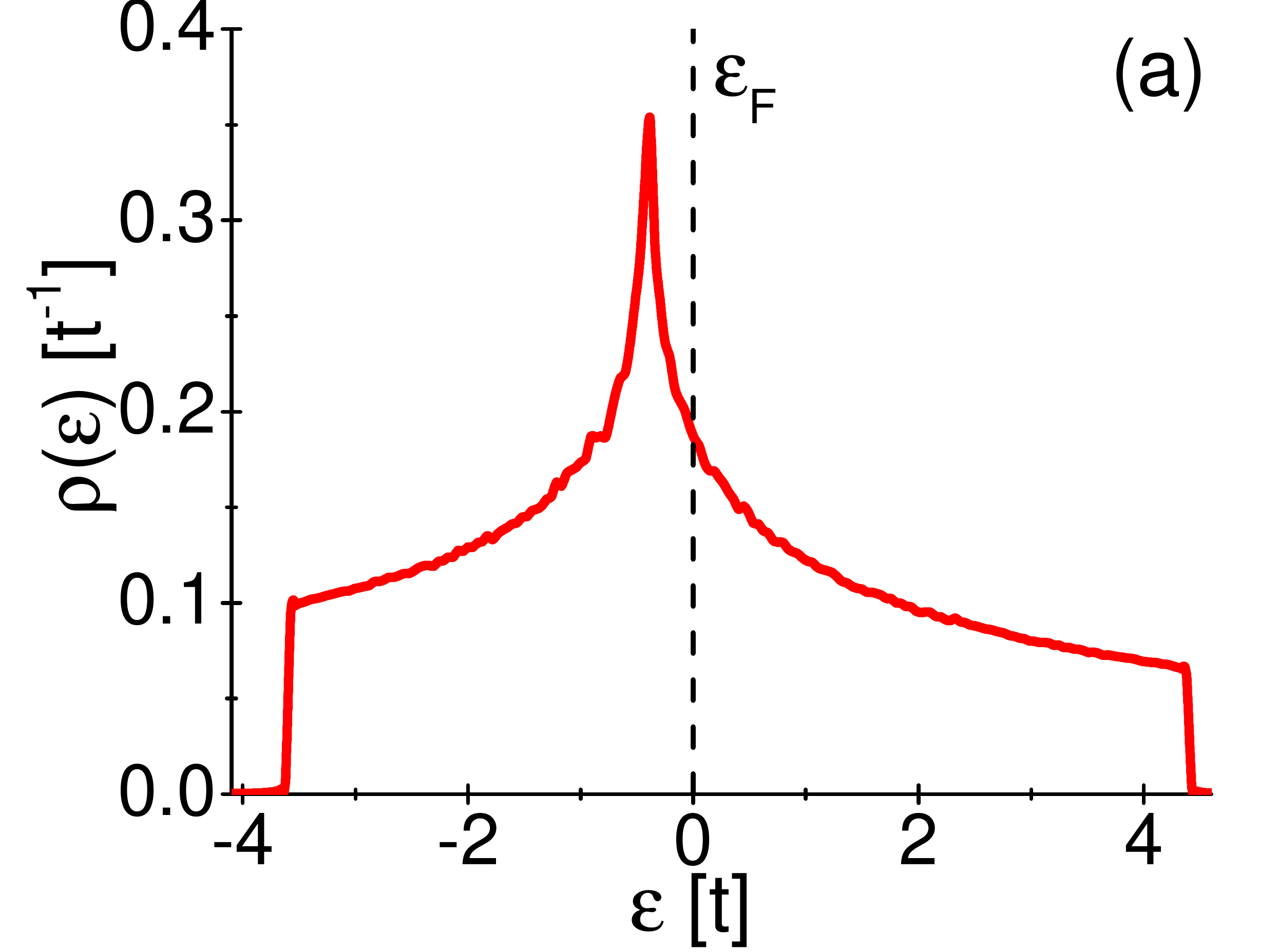}
\includegraphics[width=0.3\columnwidth]{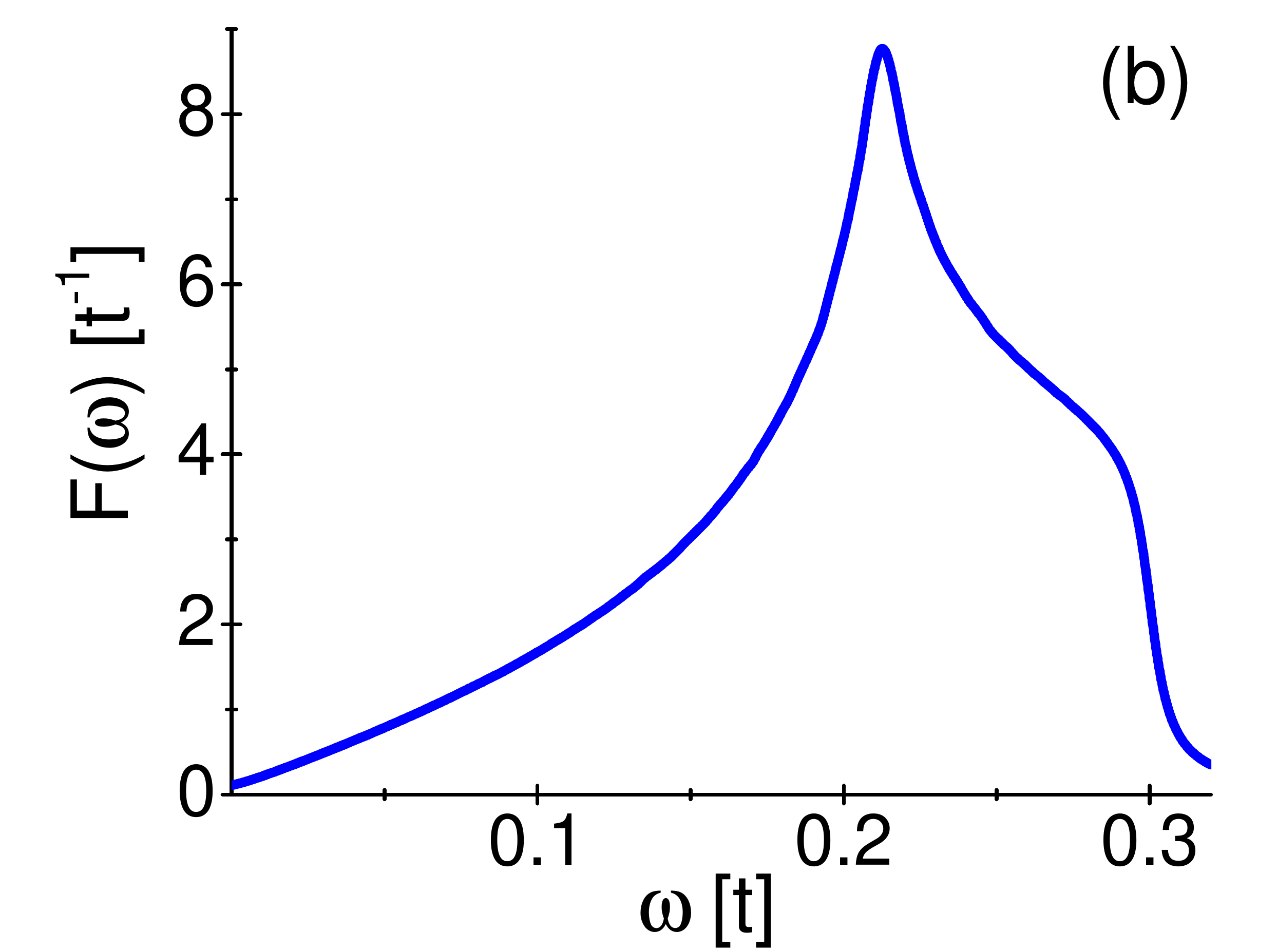}
\includegraphics[width=0.3\columnwidth]{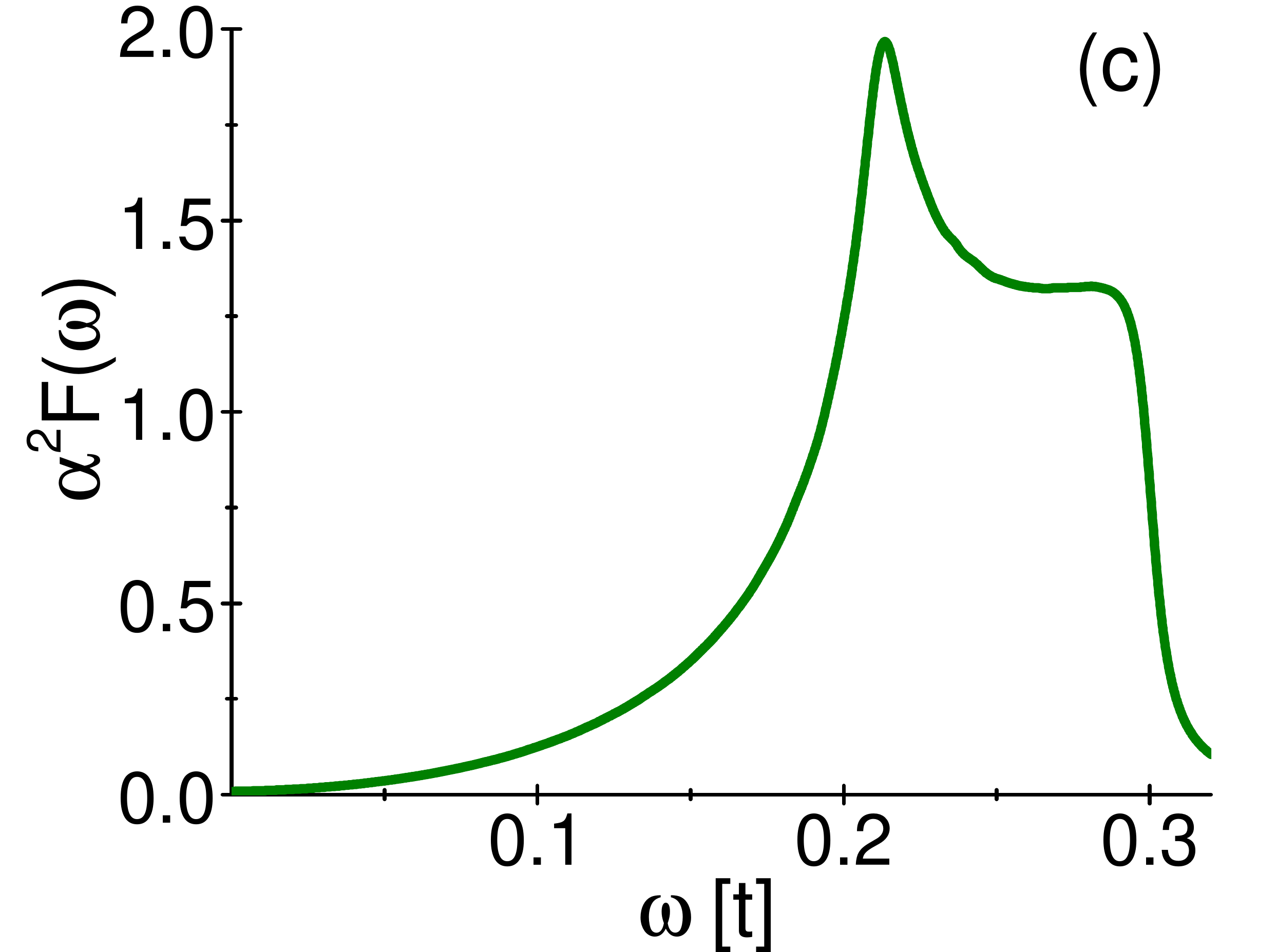}
\caption
{(a) Electron density of states function,
         (b) phonon density of states function, and (c) isotropic Eliashberg function}
\label{f00}         
\end{figure}

As the first step, let us discuss the properties of the superconducting state in the isotropic approximation and find the thermodynamic properties of the superconducting phase by means of analitical formulae.
%GP \textcolor{ForestGreen}
The obtained results shall also make us sure that the initial values of parameters ($\omega_{0}$ and $g_{0}$) were chosen in the physically acceptable way. 

The electron density of states function is described by the formula:
\begin{eqnarray} 
\rho\left(\varepsilon\right)=\frac{1}{N}\sum_{k_{x},k_{y}}\delta\left(\varepsilon-\varepsilon_{k_{x},k_{y}}\right)
=
\rho_{0}\int^{\pi}_{-\pi}dk_{x}\int^{\pi}_{-\pi}dk_{y}
\delta\left(\varepsilon-\varepsilon_{k_{x},k_{y}}\right).
\label{r04}
\end{eqnarray}
The symbol $\rho_{0}$ denotes the normalisation constant, which can be estimated from the condition: 
$\int^{W_{u}}_{W_{d}}d\varepsilon\rho\left(\varepsilon\right)=1$, where $W_{d}$ and $W_{u}$ are the upper and the lower limits of the electron band, equal to $W_{u}=4.6$~$t$ and $W_{d}=-4.1$~$t$, respectively. Additionally, $\rho_{0}=0.0254144$.
The profile of the electron density of states function is shown in \fig{f00}~(a). 
The value of the considered function at the Fermi level is relatively high ($\rho\left(0\right)=0.18724$~$t^{-1}$). 
This results from the presence of the van Hove singularity in the electron density of states \cite{VanHove1953A, Goicochea1994A, Szczesniak2002A}. 
It is relatively close to the Fermi energy level due to the fact that the hopping integral $t$ is significantly greater than $t'$ (the slight violation of the particle--hole interaction).
%
%GP \textcolor{blue}
The phonon density of states is calculated in the similar manner:
\begin{eqnarray} 
F\left(\omega\right)=\frac{1}{N}\sum_{q_{x},q_{y}}\delta\left(\omega-\omega_{q_{x},q_{y}}\right)
=
F_{0}\int^{\pi}_{-\pi}dq_{x}\int^{\pi}_{-\pi}dq_{y}\delta\left(\omega-\omega_{q_{x},q_{y}}\right).
\label{r05}
\end{eqnarray}
The $F_{0}$ value was determined from the normalisation condition: 
$\int^{\omega_{D}}_{0}d\omega F\left(\omega\right)=1$ and found to be $F_{0}=0.026079$.
The phonon density of states function is plotted in \fig{f00}~(b). 
The maximum value of the function $F\left(\omega\right)$ corresponding to the energy $\omega\sim 0.21$~$t$ is associated with the flattening of the phonon dispersion relation. For $\omega\rightarrow 0$ we obtained $F\left(\omega\right)\rightarrow 0$. As expected, the phonon density of states function vanishes for the frequency $\omega_{D}$.
%
%GP \textcolor{blue}
In the case of the Eliashberg function, we started from the general formula:
\begin{equation} 
\alpha^{2}F\left({\bf k},{\bf k}',\omega\right)=\rho\left(0\right)\sum_{\qvec} |g\left({\bf q},{\bf k}, {\bf k}'\right)|^{2}
\delta\left(\omega-\omega_{\qvec}\right).
\label{r06}
\end{equation}
Then the spectral function $\alpha^{2}F\left({\bf k},{\bf k}',\omega\right)$ was averaged over the Fermi surface in order to achieve the isotropic Eliashberg function:
\begin{equation} 
\alpha^{2}F\left(\omega\right)=\frac{1}{N^{2}}\sum_{\kvec,\kvec'}w_{\kvec}w_{\kvec '}\alpha^{2}F\left({\bf k},{\bf k}',\omega\right),
\label{r07}
\end{equation}
where: $w_{\kvec}=\delta\left(\varepsilon_{\kvec}\right)/\rho\left(0\right)$. Simple calculations led to the result:
\begin{equation} 
\alpha^{2}F\left(\omega\right)=\rho\left(0\right)\sum_{\qvec}g^{2}_{\qvec}\delta\left(\omega-\omega_{\qvec}\right).
\label{r08}
\end{equation}
We presented the isotropic Eliashberg function in \fig{f00}~(c). While comparing its profile with the profile of the phonon density of states function, one can notice a difference in their shapes. The difference characterises the contribution from the squares of the electron--phonon coupling and the electron density of states functions.

%GP \textcolor{blue}
We assessed the walue of the critical temperature from the Allen--Dynes formula \cite{Allen1975A}:
\begin{equation}
\label{r09}
k_{B}T_{C}=f_{1}f_{2}\frac{\omega_{{\rm ln}}}{1.2}\exp\left[\frac{-1.04\left(1+\lambda\right)}{\lambda}\right],
\end{equation}
however we did not take into account the depairing electron correlations modeled by Coulomb pseudopotential $\mu^{\star}$ \cite{Morel1962A} in our analysis. This correlations, if included, shift the critical temperature value down in the considered model, but they usually do not extinguish totally the superconducting phase, because the pseudopotential does not take very high values ($\mu^{\star}\sim 0.1$). The symbols used in \eq{r09} are defined in \tab{t01}.
\begin{table}
\caption{\label{t01}
%GP \textcolor{blue}
{Parameters $\lambda$, $\omega_{{\rm ln}}$, and $\omega_{2}$ represent the electron--phonon coupling constant, the logarithmic phonon frequency, and the second moment of the normalized weight function, respectively. Functions $f_{1}$ and $f_{2}$ are the strong--coupling correction function and the shape correction function, respectively.}}
 \begin{tabularx}{200pt}{cc}
Quantity & Value \\
\hline
 & \\
${\lambda=2\int^{+\infty}_0 d\Omega \frac{\alpha^2\left(\Omega\right)F\left(\Omega\right)}{\Omega}}$ & $2.06$\\
$\omega_{{\rm ln}}=\exp\left[\frac{2}{\lambda}
\int^{+\infty}_{0}d\Omega\frac{\alpha^{2}F\left(\Omega\right)}
{\Omega}\ln\left(\Omega\right)\right]$ & $0.109$~$t$\\
 & \\ 
$\sqrt{\omega_{2}}= 
\left[\frac{2}{\lambda}
\int^{+\infty}_{0}d\Omega\alpha^{2}F\left(\Omega\right)\Omega\right]^{1/2}$ & $0.208$~$t$\\
& \\
$f_{1}=\left[1+\left(\frac{\lambda}{\Lambda_{1}}\right)^{\frac{3}{2}}\right]^{\frac{1}{3}}$ & - \\
 & \\
$f_{2}= 1+\frac
{\left(\frac{\sqrt{\omega_{2}}}{\omega_{\rm{ln}}}-1\right)\lambda^{2}}
{\lambda^{2}+\Lambda^{2}_{2}}$ & - \\
 & \\
$\Lambda_{1}= 2.46$ & - \\
 & \\
$\Lambda_{2}=1.82\left(\sqrt{\omega_{2}}/\omega_{\ln}\right)$ & - \\
 & \\
\end{tabularx}
\end{table}
%
%GP \textcolor{blue}
The dimensionless coefficients corresponding to those used in the BCS theory can be also calculated in a relatively easy way \cite{Bardeen1957A, Bardeen1957B}:
\begin{eqnarray}
\label{r10} 
R_{\Delta}&=&\frac{2\Delta(0)}{k_{B}T_{C}},\\ 
R_{C}&=&\frac{\Delta C\left(T_{C}\right)}{C^{N}\left(T_{C}\right)},\\ 
R_{H}&=&\frac{T_{C}C^{N}\left(T_{C}\right)}{H_{C}^{2}\left(0\right)}. 
\end{eqnarray}
Values of these coefficients calculated within the BCS theory are equal to $3.53$, $1.43$, and $0.168$, respectively. In the considered case, however, when we can use the explicit form of the Eliashberg function, the coefficients $R_{\Delta}$, $R_{C}$ and $R_{H}$ should be assessed on the basis of the formulae \cite{Carbotte1990A}:
\begin{eqnarray}
\label{r11} 
R_{\Delta}&=&3.53\left[1+12.5\left[\frac{T_{C}}{\omega_{\rm ln}}\right]^{2}\ln\left(\frac{\omega_{\rm ln}}{2T_{C}}\right)\right],\\ 
R_{C}&=&1.43\left[1+53\left[\frac{T_{C}}{\omega_{\rm ln}}\right]^{2}\ln\left(\frac{\omega_{\rm ln}}{3T_{C}}\right)\right],\\ 
R_{H}&=&0.168\left[1-12.2\left[\frac{T_{C}}{\omega_{\rm ln}}\right]^{2}\ln\left(\frac{\omega_{\rm ln}}{3T_{C}}\right)\right]. 
\end{eqnarray}
For the values of initial parameters assumed in the work we obtained: $k_{B}T_{C}=0.029$~$t$. 
Additionally: $R_{\Delta}=5.5$, $R_{C}=2.64$ oraz $R_{H}=0.135$. 
It means that the physical system is characterised by a strong electron--phonon coupling, what can be also concluded from the value of $\lambda$ (see \tab{t01}). If $t=250$~meV, then the critical temperature is equal to $84.19$~K.

\begin{figure}
\includegraphics[width=0.5\columnwidth]{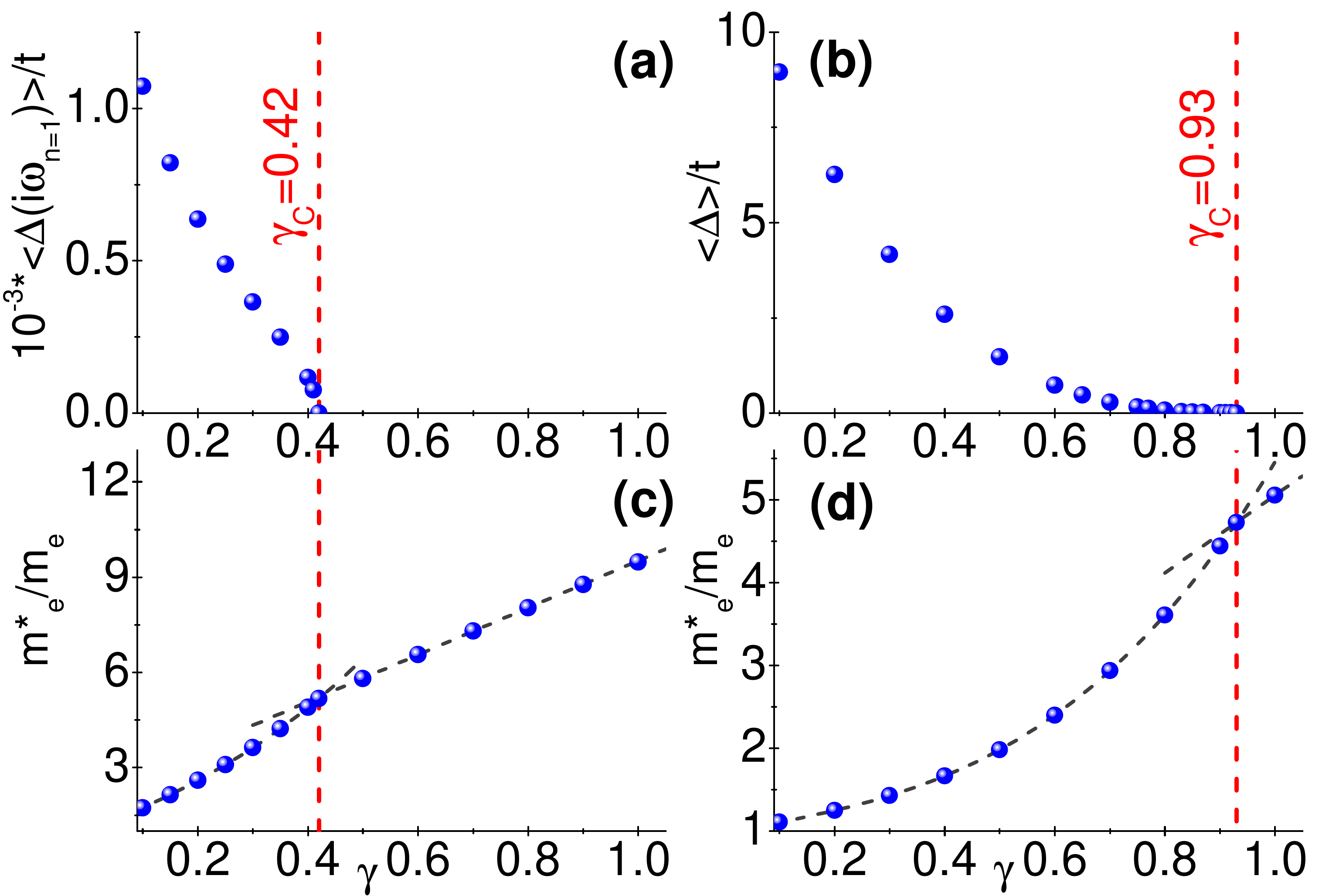}
\caption{(a) The dependence of the average value of the order parameter $\left<\Delta(i\omega_{n=1})\right>$ on $\gamma$ 
             in the scheme of full self--consistency of the Eliashberg equations.  
         (b) The same dependence as in (a) -- results of the momentum approximation, within which the self--consistency with respect 
             to Matsubara frequency is omitted: 
             $\Delta_{\bf k}\left(i\omega_{n}\right)\sim\Delta_{\bf k}$ and $Z_{\bf k}\left(i\omega_{n}\right)\sim Z_{\bf k}$.
         (c) The results concerning the effective mass of the electron achieved in the scheme of full self--consistency. 
             It should be noticed that a characteristic deflection occurs in the profile of the effective electron mass versus 
             the unbalance parameter curve for $\gamma=\gamma_{C}$. It results from the disappearance of the superconducting state.
         (d) The same dependence as in (c) -- the scheme of the momentum approximation.         
             We assumed $k_{B}T_{0}=0.0001$~$t$ for the purpose of numerical calculations.}
\label{f01}         
\end{figure}
\begin{figure}
\includegraphics[width=1\columnwidth]{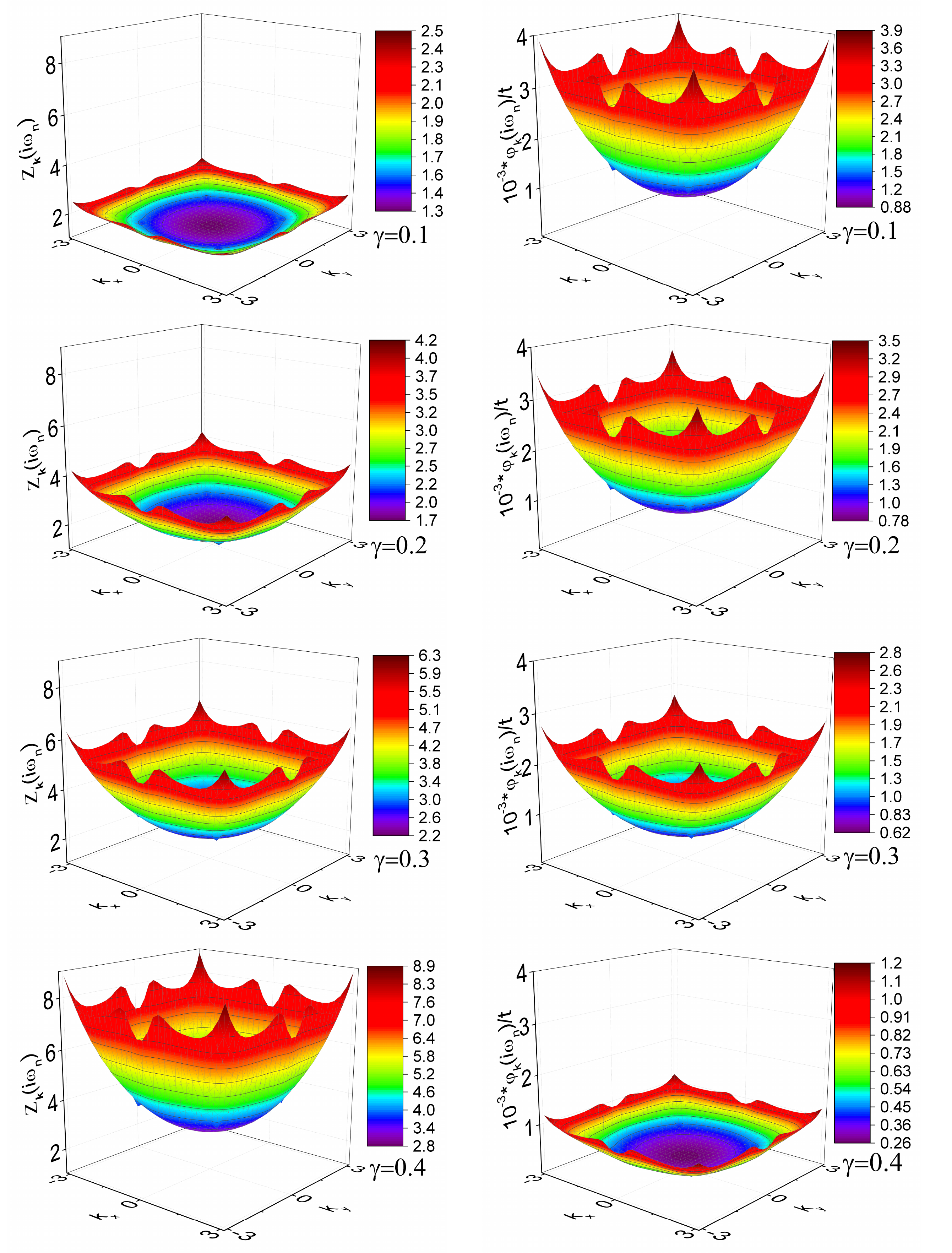}
\caption{The dependence of the wave function renormalising factor and of the order parameter function on the wave vector for selected values of the unbalance parameter. The results obtained for the scheme of fully self--consistent solution of the Eliashberg equations 
($k_{B}T_{0}=0.0001$~$t$).}
\label{f02}         
\end{figure}
\begin{figure}
\includegraphics[width=\columnwidth]{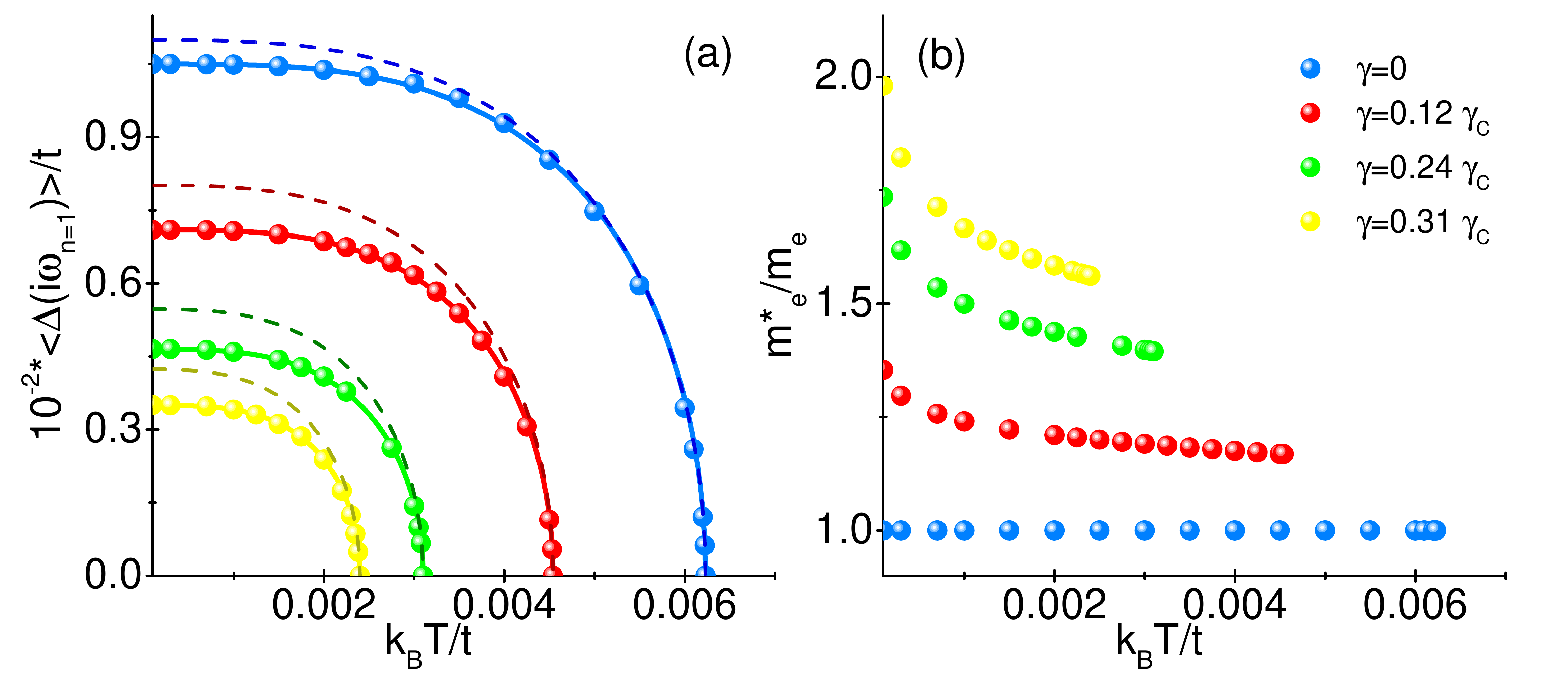}
\caption{(a) The influence of temperature on the averaged value of the order parameter $\left<\Delta(i\omega_{n=1})\right>$. 
             %GP \textcolor{blue}
             Numerical results obtained for the unbalance parameter equal to $0$, $0.12\gamma_{C}$, $0.24\gamma_{C}$, and $0.31\gamma_{C}$.             
             The continuous curves were found by means of the formula: 
             $\Delta\left(T\right)=\Delta\left(0\right)\sqrt{1-\left(T/T_{C}\right)^{\Gamma}}$, where 
             $\Gamma\in\{3.35, 3.3, 3.3, 3.4\}$ for $\gamma\in\{0, 0.12\gamma_{C},0.24\gamma_{C},0.31\gamma_{C}\}$, respectively. 
             Additionally \mbox{$\Delta\left(0\right)=\Delta\left(T_{0}\right)$}.
             For the BCS-type curves (dashed lines), it should be assumed that $\Gamma=3.0$ and $2\Delta\left(0\right)/k_{B}T=3.53$ 
             \cite{Bardeen1957A, Bardeen1957B, Eschrig2001A}.
         (b) The temperature dependence of the effective electron mass.
        }
\label{f03}         
\end{figure}

%GP \textcolor{blue}
Let us proceed to the analysis of the thermodynamic properties of the superconducting state using the full Eliashberg equations. It occurs that the achieved results are significantly different from the results obtained within the isotropic approximation.

The dependence between the averaged value of the order parameter $\left<\Delta(i\omega_{n=1})\right>=\frac{1}{N}\sum_{\kvec}\Delta_{\kvec}(i\omega_{n=1})$ and the value of the unbalance parameter is plotted in \fig{f01}~(a). It can be easily noticed that the superconducting phase ceases to exist for $\gamma\geq\gamma_{C}=0.42$. Let us remember that within the momentum approximation with a constant value of the electron--phonon coupling function 
($g_{\qvec}\sim g$) we got $\gamma_{C}=0.94$ \cite{Szewczyk2018A}. A similar value, $\gamma_{C}=0.93$, is obtained from calculations in which the momentum transfer is explicitly taken into account in the electron--phonon coupling function (see \fig{f01}~(b)). 
%GP \textcolor{ForestGreen}
It should be strongly emphasized that our result does not mean that the balanced phonon--induced superconducting state cannot exists on a square lattice in general. Attention should be paid to the fact that, despite the relatively sophisticated analysis presented in the work, we applied also a series of approximations. They can significantly influence the final results. We would like to mention here, in particular, the lack of explicit dependence of the electron--phonon coupling function on the $\bf k$ vector, omitting the vertex corrections to the electron--phonon interaction in the analysis (and yet $\omega_{D}/W\sim 0.1$, where $W\sim 4$~$t$), or disregarding the anharmonic corrections,~etc. Therefore our result should be considered as physically interesting, but acquired under some specific assumptions. Formulation of general conclusions with respect to the phonon--induced superconducting state on a square lattice demands for still deeper analysis in future.

The dependence of the electron effective mass ($m^{\star}_{e}$) on the value of the unbalance parameter $\gamma$ is presented in \mbox{\fig{f01}~(c)}, wherein $m^{\star}_{e}=\left<Z(i\omega_{n=1})\right>m_{e}$, and the symbol $m_{e}$ denotes the band mass of an electron. 
%GP \textcolor{blue}
The obtained results prove that the superconducting phase on a square lattice ceases due to the fact that the electron effective mass takes high values ($\left[m^{\star}_{e}\right]_{\gamma=\gamma_{C}}=5.18$~$m_{e}$ and 
%GP \textcolor{red}
$\left[m^{\star}_{e}\right]_{\gamma=1}=9.48$~$m_{e}$), 
%GP \textcolor{blue}
what results in the decrease in the critical temperature: 
$T_{C}\sim \exp\left(-m^{\star}_{e}\right)$ \cite{McMillan1968A, Allen1975A}. The values of the electron effective mass for the momentum approximation and the explicitly given $g_{\bf q}$ are significantly less (\fig{f01}~(d)). 
%GP \textcolor{red}
Quite similar values of the electron effective mass we can observe for the momentum approximation ($\left[m^{\star}_{e}\right]_{\gamma=\gamma_{C}}\sim 4.73$~$m_{e}$ and 
%GP \textcolor{red}
$\left[m^{\star}_{e}\right]_{\gamma=1}=5.46$~$m_{e}$) under the additional assumption $g_{\qvec}\sim g$ \cite{Szewczyk2018A}.
%GP \textcolor{ForestGreen}
Let us notice that the above mentioned results agree with the fundamental result achieved by Cooper, who proved that the Fermi system is unstable towards the formation of bound Cooper pairs for any infinitesimally small coupling \cite{Cooper1956A}. Cooper's considerations dealt with the case of $\gamma=0$ ($m^{\star}_{e}=m_{e}$), for which we also observed an induction of the superconducting condensate. 

As far as the results achieved within the isotropic approximation ($\Delta_{\bf k}\left(i\omega_{n}\right)\sim\Delta\left(i\omega_{n}\right)$ and \mbox{$Z_{\bf k}\left(i\omega_{n}\right)\sim Z\left(i\omega_{n}\right)$)} are concerned, 
the unbalanced Eliashberg equations were solved by Cappelluti and Ummarino in \cite{Cappelluti2007A}. For the case of 3D system, the existence of the balanced superconducting state was confirmed, and even the unbalanced superconducting state of the $\gamma$ parameter value significantly exceeding unity was stated. The calculations which we performed for the isotropic approximation on a square lattice also predict induction of either the balanced or the unbalanced superconducting state for $\gamma>1$ (see Appendix \ref{D01}). 
The above results mean that the isotropic approximation, at least with respect to the square lattice, 
%GP \textcolor{ForestGreen}
should be applied with considerable caution.  

Having acquired the explicit solutions of the Eliashberg equations, one can determine the regions of the first Brillouin zone, for which the function $m^{\star}_{e}\left({\bf k}\right)$ contributes particularly greatly to the averaged value of the effective electron mass. We present diagrams showing the values of the wave function renormalisation factor versus the wave vector in \fig{f02}~(left column). It can be seen that an increase in the effective electron mass is generated mainly by increase in the value of $m^{\star}_{e}\left({\bf k}\right)$ function in the whole Brillouin zone. Although at the 
boundaries of the Brillouin zone it is particularly high. The order parameter function $\varphi_{\bf k}\left(i\omega _{n=1}\right)$, which is plotted in \fig{f02}~(right column), has a similar form.  
\begin{figure}
\includegraphics[width=0.5\columnwidth]{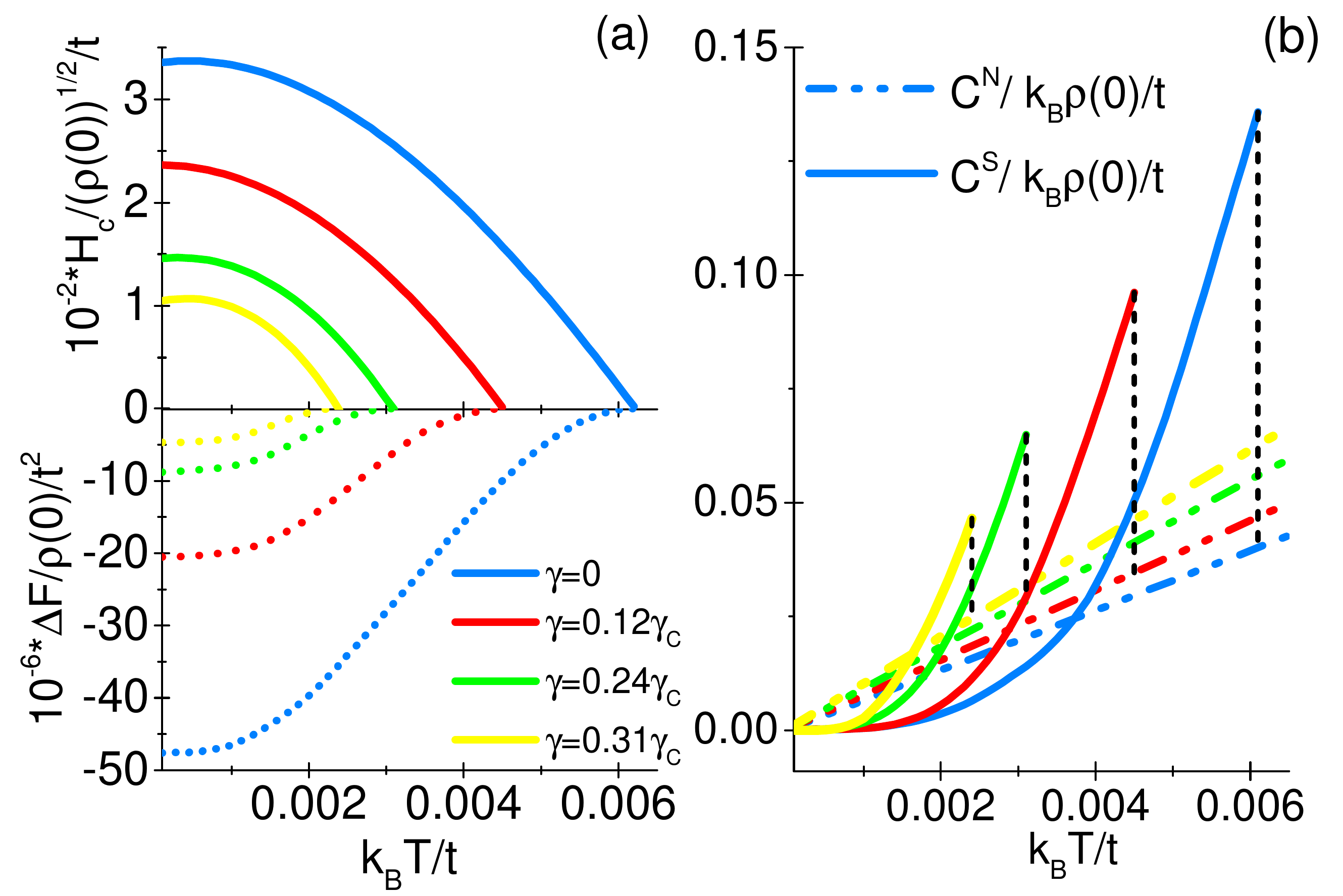}
\caption{(a) The difference in free energy between the superconducting and the normal state (lower curve) 
             and the thermodynamic critical field (upper curve). 
         (b) The specific heat for the superconducting and the normal state. The black vertical line indicates the specific heat jump in the critical 
             temperature.}
\label{f04}         
\end{figure}

Our results prove that only the unbalanced superconductiong state of $\gamma<\gamma_{C}=0.42$ can be induced on a~square lattice. 
%GP \textcolor{blue}
{We selected four values of the unbalance parameter ($0$, $0.12\gamma_{C}$, $0.24\gamma_{C}$, and $0.31\gamma_{C}$) in order to analyse thermodynamical properties of this state.} The achieved results are plotted in \fig{f03} for both the order parameter and the wave function renormalising factor. It can be seen that in every case the dependence of the order parameter on temperature deviates considerably from the one predicted by the BCS theory \cite{Bardeen1957A, Bardeen1957B}. 
%GP \textcolor{blue}
{Taking into account the ratio $R_{\Delta}=2\Delta\left(0\right)/k_BT_C$ we get:
$R^{0}_{\Delta}=3.37$, $R^{0.12\gamma_{C}}_{\Delta}=3.13$, $R^{0.24\gamma_{C}}_{\Delta}=3$, and $R^{0.31\gamma_{C}}_{\Delta}=2.92$.} 
Please notice that the BCS theory predicts $R_{\Delta}= 3.53$. 

The influence of temperature on the value of the effective electron mass is presented in \fig{f03}~(b). 
%GP \textcolor{blue}
{The dependence of $m^{\star}_{e}$ on $T$ is weak, as it is both in the isotropic case and in the momentum approximation \cite{Szczesniak2006B, Szewczyk2018A}.}
  
The thermodynamics of the superconducting state is fully determined by the values of both the order parameter and the wave function renormalising factor. Considering the averaged values of the discussed functions, we calculated the difference in free energy between the superconducting and the normal state \cite{Bardeen1964A}:
\mbox{$\Delta F/\rho\left(0\right)=-\left(2\pi/\beta\right)\sum^{M}_{n=1}[\sqrt{\omega^2_n+\left<\Delta_n\right>^2}-|\omega_n|]
[\left<Z^{\left(S\right)}_n\right>-\left<Z^{\left(N\right)}_n\right>\frac{|\omega_n|}{\sqrt{\omega^2_n+\left<\Delta_n\right>^2}}]$}. 
Symbols $Z^{\left(S\right)}_n$ and $Z^{\left(N\right)}_n$ denote the values of the wave function renormalising factor in the superconducting and the normal state, respectively. The thermodynamic critical field should be determined from the formula: $H_{C}=\sqrt{-8\pi\Delta F}$, whereas the difference in the specific heat between the superconducting and the normal state can be found using the formula: 
$\Delta C=C^{S}-C^{N}=-T\frac{d^2\Delta F}{dT^2}$, where: $C^{N}=\gamma{T}$. The Sommerfeld constant is equal to: 
$\gamma=\frac{2}{3}\pi^{2}k_{B}^{2}\rho(0)\left(1+\lambda\right)$. The achieved results we showed in \fig{f04}. Then we calculated the values of the non-dimensional thermodynamic ratios: 
$R_C=\left[C^{S}\left(T_C\right)-C^{N}\left(T_C\right)\right]/C^N\left(T_C\right)$ and 
$R_H=T_CC^N\left(T_C\right)/H_C^2\left(0\right)$. The considered results correspond to the universal constants in the BCS theory, the latter taking the values equal to $1.43$ and $0.168$, respectively \cite{Bardeen1957A, Bardeen1957B}. 
%GP \textcolor{blue}
{For the superconducting state on a~square lattice, we arrived at the results which differ significantly from the ones occurring 
in the BCS theory: 
$R^{0}_{C}=2.4$, $R^{0}_{H}=0.227$; 
$R^{0.12\gamma_{C}}_{C}=1.89$, $R^{0.12\gamma_{C}}_{H}=0.283$;
$R^{0.24\gamma_{C}}_{C}=1.28$, $R^{0.24\gamma_{C}}_{H}=0.413$;
$R^{0.31\gamma_{C}}_{C}=0.89$, $R^{0.31\gamma_{C}}_{H}=0.595$.}

To summarize, we proved that -- under the assumptions made with respect to the electron--phonon coupling function, vertex corrections, and anharmonic effects -- the balanced phonon--induced superconducting state cannot be generated on a square lattice. On the other hand, the linear electron--phonon interaction can induce the unbalanced superconducting state in cases for which the unbalance parameter takes a value less than $0.42$. 
%GP \textcolor{ForestGreen}
This value is much lower than $\gamma_{C}$ estimated in the static boundary ($0.93$). It means that the dynamic effects modeled by the explicit dependence of the order parameter and the wave function renormalising factor on the Matsubara frequency are very important for the superconducting state on a square lattice. This result matched our expectations, given that the interaction between electrons is transmitted via slow phonons. 
%GP \textcolor{blue}
The reason for a decrease in the critical temperature value with an increase in 
$\gamma$ parameter value is the anomalously high increase in the effective electron mass. 
%GP \textcolor{ForestGreen}
It should be rightly stressed that the possibility of induction of the superconducting state on a square lattice most probably depends strongly on the form of the electron (phonon) dispersion relation, the electron--phonon coupling function, or the assumed approximations. Therefore one cannot claim that the electron--phonon interaction in low--dimensional systems (2D) is unimportant with respect to the superconducting state. Our result evidence that the analysis of the thermodynamical properties of the superconducting phase, if too cursory, e.g. within the isotropic approximation, can generate false results. 
%GP \textcolor{ForestGreen}
This conclusion is of particular importance in the case of the most recent low--dimensional systems (decorated graphene \cite{Profeta2012A, Pesic2014A, Guzman2014A, Kaloni2013A}, silicene \cite{Wan2013A} or phosphorene \cite{Ge2015A, Shao2014A}), in which the superconducting state can be induced.  

The performed numerical analysis proved that the unbalanced superconducting state is described by the thermodynamic parameters, which values differ significantly from the values predicted by the BCS theory. 

Please notice that our results can make the understanding of the pairing mechanism in cuprates considerably easier \cite{Dagotto1994A, Szczesniak2012A}. It is generally accepted fact that electrons in cuprates form a strongly correlated system \cite{Emery1987A, Littlewood1989A, Hybertsen1990A}. Numerical calculations carried out for the Hubbard model demonstrate that the value of the on-site $U$~integral is equal to about $5$~eV \cite{Hybertsen1990A}. It should be strongly emphasised that the balanced linear electron--phonon interaction is too weak to induce the experimentally observed superconducting state (the full discussion of the considered problem at the {\it ab initio} level can be found in \cite{Bohnen2003A}). On the other hand, there exists a good deal of experimental data which indicate that the interaction between electrons and phonons in cuprates is really significant. Let us recall the results obtained by means of the ARPES method, which demonstrated the existence of a break in the energy spectrum near the phonon energy \cite{Damascelli2003A, Cuk2005A}. Additionally, the ARPES method made possible to discover the isotope effect of the real part of the self--energy \cite{Gweon2004A}. The isotope effect is also related to the critical temperature, what is particularly distinct in the strongly underdoped regions \cite{Franck1994A}. Moreover, the vibrations of the crystal lattice modify the penetration depth and the results of Raman measurements \cite{Hofer2000A, Schneider2005A}.

Therefore an obvious suggestion comes into mind that the high--temperature superconducting state in cuprates may be induced by the strongly unbalanced electron--phonon interaction. This suggestion is quite probable for the reason that it is very difficult to prove the existence of the superconducting state of sufficiently high critical temperature value in pure electron models with $U>0$. Such a~problem, however, does not occur in effective models with $U<0$. Our hypothesis is also confirmed by the results achieved by Kim and Tesanovic \cite{Kim1993A}, who showed, on the example of ${\rm La_{2-x}Sr_{x}CuO_{4}}$, ${\rm (Y_{1-x}Pr_{x})Ba_{2}Cu_{3}O_{7-y}}$ and ${\rm YBa_{2-x}La_{x}Cu_{3}O_{7}}$ compounds, that the strong Coulomb correlations do not suppress the phonon pairing mechanism over a wide doping range. 

%%%%%%%%%%%%%%%%%%%%%%%%%%%%%%%%%%%%%%%%%%%%%%%%%%%%%%%%%%%%%%%%%%%%%%%%%%%%%%%%%%%%%%%%%%%%%%%%%%%%%%%%%%%%%%%%%%%%%%%%%%%%%%%%%%%%%%%%%%%%%%%%%%%%%%%%%%
\appendix

%%%%%%%%%%%%%%%%%%%%%%%%%%%%%%%%%%(A)
\section{\label{D01} Dependence of the order parameter value on the number of momentum lattice points and the number of Matsubara frequencies}

\begin{figure}[h]
\includegraphics[width=0.5\columnwidth]{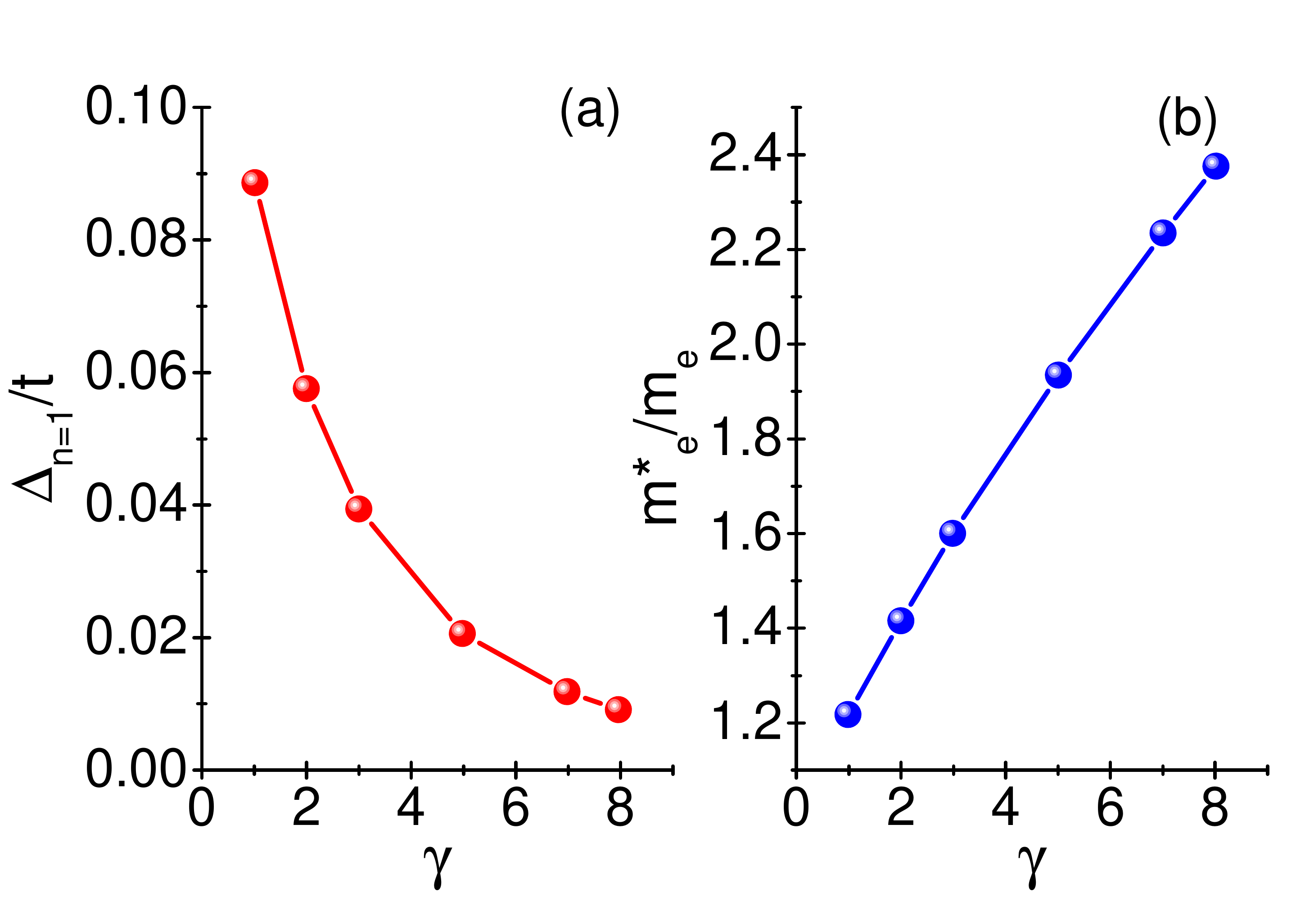}
\caption{
%GP \textcolor{blue}
{
(a) The influence of change in the number of the momentum lattice points $N$ on the averaged value of the order parameter at constant number of Matsubara frequencies $n=200$.  
(b) The dependence of the averaged value of the order parameter on the number of Matsubara frequencies $n$ for constant number of momentum lattice points $N=200$.}
        }
\label{f05}         
\end{figure}

%GP \textcolor{blue}
We checked the number of momentum lattice points and the number of Matsubara frequencies for which the value of the order parameter is saturated in the temperature $k_{B}T_{0}=0.0001$~$t$. We plotted the results in \fig{f05}. It can be assumed with good accuracy that the averaged value of the order parameter is saturated for $N\sim 79$ and $n\sim 67$, under the assumption that $n=200$ and $N=200$, respectively. 
It is worth remembering that the value of the order parameter will be saturated much faster for higher temperature values.

%

%%%%%%%%%%%%%%%%%%%%%%%%%%%%%%%%
\section{\label{D02} The unbalanced superconducting state on a square lattice -- the isotropic approximation}

\begin{figure}
\includegraphics[width=0.5\columnwidth]{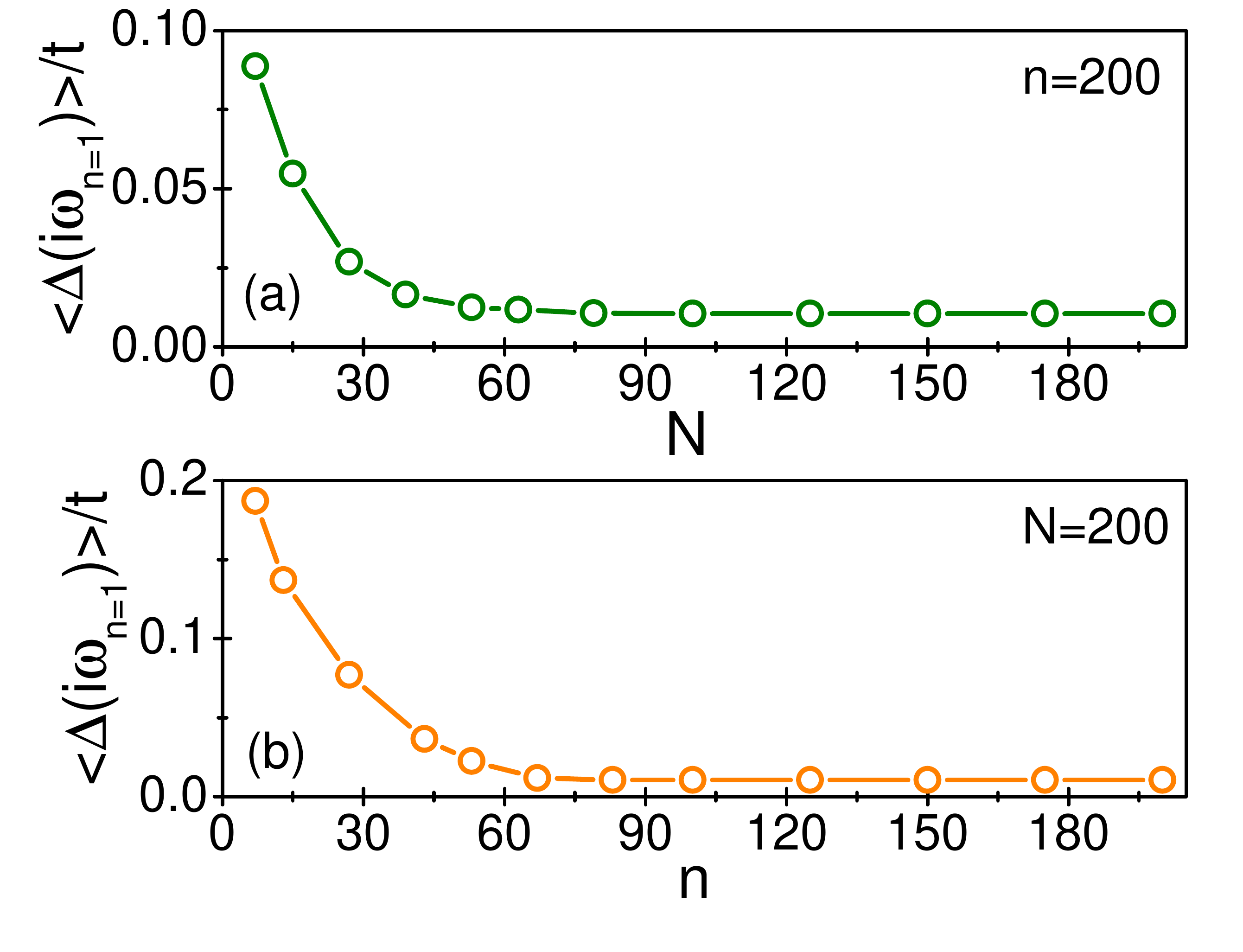}
\caption{The dependence of the order parameter (a) and of the effective electron mass (b) on the value of $\gamma$ parameter.
         Assumed values: $\lambda=2t$, $\omega_{D}=0.3t$, and 
%GP \textcolor{blue}
{$k_{B}T_{0}=0.0001t$}. }
\label{f06}         
\end{figure}

Let us discuss the issue of the unbalanced superconducting state within the isotropic approximation. The Eliashberg equations take the form: 
\begin{eqnarray}
\label{rB1}
\varphi_{n}&=&\frac{1}{\beta}\sum_{m}K'\left(n-m\right)\varphi_{m}\sum_{\kvec}p_{\kvec}^{-1}\left(m\right),\\  
Z_{n}&=&1+\frac{\gamma}{\omega_{n}\beta}\sum_{m}K'\left(n-m\right)Z_{m}\omega_{m}\sum_{\kvec}p_{\kvec}^{-1}\left(m\right),\nonumber  
\end{eqnarray}
where: $p_{\kvec}\left( m\right)=\left(Z_m\omega_m\right)^2+\varepsilon^{2}_{\kvec}+\varphi_m^2$. 
The pairing kernel of the electron--phonon interaction is given by: \mbox{$K'\left(n-m\right)=\lambda\frac{\nu^2}{\left(n-m\right)^2+\nu^2}$}, where 
$\lambda$ is the electron--phonon coupling constant. We used Kresin's method of introducing the average phonon frequency: 
$\left<\Omega \right>\sim\omega_{D}$. In this case: $\nu=\beta\omega_D/2\pi$ \cite{Kresin1984A, Kresin1987A}. 
The density of states per spin direction $\rho\left(\varepsilon\right)$ is reproduced by: 
$\rho\left(\varepsilon\right)=b_1\ln\left|\varepsilon/b_2\right|$, where $b_{1}=-0.04687t^{-1}$ and $b_{2}=21.17796t$ \cite{Szczesniak2006B}. Hence we obtain: $\sum_{\kvec}p_{\kvec}^{-1}\left(m\right)\simeq \int^{W}_{-W}d\varepsilon\rho\left(\varepsilon\right)p_{\varepsilon}^{-1}\left(m\right)$, 
where $W=4t$ is half of the band width.
 
The dependence of the order parameter on $\gamma$ is plotted in \fig{f06}~(a). It can be seen that, in the case of the isotropic approximation, the superconducting state on the square lattice is induced even for high values of the unbalance parameter.
The results concerning the effective electron mass are presented in \fig{f06}~(b). 
%GP \textcolor{blue}
{The values of $m^{\star}_{e}$ in the isotropic approximation are not very high, as compared with the accurate results (see \fig{f01}~(c)).} 

%%%%%%%%%%%%%%%%%%%%%%%%%%%%%%%%%%%%%%%%%%%%%%%%%%%%%%%%%%%%%%%%%%%%%%%%%%%%%%%%%%%%%%%%%%%%%%%%%%%%%%%%%%%%%%%%%%%%%%%%%%%%%%%%%%%%%%%%%%%%%%%%%%%%%%%%%%
%
\bibliography{Bibliography}
\end{document}